\documentclass[twocolumn,showpacs,aps,prl,superscriptaddress,letterpaper,floatfix]{revtex4-1}
\usepackage{graphicx}
\usepackage{dcolumn}
\usepackage{amsmath}
\usepackage[dvips]{color}
\usepackage{bm}

\RequirePackage{xspace}

\newcommand{\gev}{\ensuremath{\mathrm{\,Ge\kern -0.1em V}}\xspace}
\newcommand{\gevcc}{\ensuremath{{\mathrm{\,Ge\kern -0.1em V\!/}c^2}}\xspace}
\newcommand{\tev}{\ensuremath{\mathrm{\,Te\kern -0.1em V}}\xspace}

\newcommand{\EPEM}{\ensuremath{e^+e^-}}

\newcommand{\GG}{\mbox{$\gamma\gamma$}}

\newcommand{\GEV}{\mbox{GeV}}

\newcommand{\CM}{\mbox{cm}}
\newcommand{\M}{\mbox{m}}
\newcommand{\MM}{\mbox{mm}}

\newcommand{\MKM}{\mbox{$\mu$m}}

\newcommand{\n}{\mbox{$n_f$}}

\newcommand{\E}{\mbox{$\epsilon$}}
\newcommand{\EN}{\mbox{$\epsilon_n$}}
\newcommand{\EI}{\mbox{$\epsilon_i$}}
\newcommand{\ENI}{\mbox{$\epsilon_{ni}$}}
\newcommand{\ENX}{\mbox{$\epsilon_{nx}$}}
\newcommand{\ENY}{\mbox{$\epsilon_{ny}$}}
\newcommand{\EX}{\mbox{$\epsilon_x$}}
\newcommand{\EY}{\mbox{$\epsilon_y$}}
\newcommand{\BI}{\mbox{$\beta_i$}}

\newcommand{\SX}{\mbox{$\sigma_x$}}
\newcommand{\SY}{\mbox{$\sigma_y$}}

\newcommand{\SI}{\mbox{$\sigma_i$}}
\newcommand{\SIP}{\mbox{$\sigma_i^{\prime}$}}
\begin{document}

\preprint{}

\title{ \boldmath Laser cooling of electron beams for linear colliders}

\author{V.~I.~Telnov \\ {\it Institute of Nuclear Physics, 630090, Novosibirsk, Russia}}


\date{October 28, 1996}

\begin{abstract}
   A novel method of electron beam cooling is considered which can be
used for linear colliders. The electron beam is cooled during
collision with focused powerful laser pulse. With reasonable laser
parameters (laser flash energy about 10 J) one can decrease transverse
beam emittances by a factor about 10 per one stage. The ultimate
transverse emittances are much below those achievable by other methods.
Beam depolarization  during  cooling is about 5--15 \% for one
stage. This method is especially useful for photon colliders and opens
new possibilities for \EPEM\ colliders.

\end{abstract}


\maketitle

To explore the energy region beyond LEP-II, linear colliders (LC) with
center--of--mass energy 0.5--2 TeV are developed now in the main
accelerator centers \cite{LOW}. Besides \EPEM\ collisions,  linear
colliders  can ``convert'' electrons to high energy photons using the
Compton backscattering of laser light, thus obtaining \GG\ and $\gamma
e$ collisions with energies and luminosities close to those in \EPEM\
collisions \cite{GKST81}-\cite{BERK}.

To attain  high luminosity, beams in linear colliders should be very
tiny. At the interaction point (IP) in the current LC
designs \cite{LOW}, beams with transverse sizes as low as \SX/\SY\
$\sim$ 200/4 nm are planned. Beams for \EPEM\ collisions should be
flat in order to reduce beamstrahlung energy loss. For \GG\
collision, the beamstrahlung radiation is absent and
beams with smaller \SX\ can be used \cite{TEL90,TEL95} to obtain higher
luminosity.

The transverse beam sizes are determined by the emittances \EX, and \EY.
The beam sizes at the interaction point (IP) are $\sigma_i=\sqrt{\EI\
\BI\ }$, where \BI\ is the beta function at the IP.  With the
increase of the beam energy  the emittance of the bunch
decreases: $\EI=\ENI/\gamma$, where $\gamma=E/mc^2, $ \ENI\ is the
{\it normalized} emittance.

The beams with a small \ENI\ are usually prepared in damping rings
which naturally produce bunches with $\ENY\ll\ENX$ \cite{WID}.  Laser
RF photoguns can also produce beams with low emittances \cite{TRAV}.
However, for linear colliders it is desirable to have smaller emittances.

In this paper, a new method of electron beam cooling is discussed which
allows further reduction of the transverse emittances after damping
rings or guns by 1--3 orders of magnitude \cite{TEL96}.

The idea of laser cooling of electron beams is very simple.(This idea was
mentioned in the talk given by B.Palmer at the Berkeley Workshop on
Gamma--Gamma colliders \cite{palmer}, but actually this method was not
studied up to now.)   During a collision with optical
laser photons (in the case of strong field it is more correct to
consider the interaction of an electron with an electromagnetic wave)
the transverse distribution of electrons ($\sigma_i$) remains almost
unchanged. Also the angular spread ($\sigma_i^{\prime}$) is almost
constant, because for photon energies (a few eV) much lower than the
electron beam energy (several GeV) the scattered photons follow the
initial electron trajectory with a small additional spread. So, the
emittance $\EI = \SI \SIP$ remains almost unchanged. At the same time,
the electron energy decreases from $E_0$ down to $E$. This means that
the transverse normalized emittances have decreased: $ \EN = \gamma \E
= \EN_0(E/E_0)$.

One can reaccelerate the electron beam up to the initial energy and
repeat the procedure. Then after N stages of cooling $ \EN /\EN _0 =
(E/E_0)^N$ (if \EN\ is far from its limit).

In this method, we have to consider first the following problems: 1)
requirements on laser parameters (these parameters should be
attainable); 2) an energy spread of the beam after cooling (at the final
energy of a linear collider it is necessary to have $\sigma_E/E \sim
0.1 \%$; also with a large energy spread it is difficult to repeat
cooling many times due to the problem of beam focusing); 3) the limit on
the final normalized emittances (it is desirable to have this limit
lower than that obtained with storage rings and photoguns);
4) depolarization of electron beams (polarization is very important for
linear colliders).

In the cooling region, a laser photon with  energy $\omega_0$ (wave
length $\lambda$) collides almost head--on with an electron of
energy $E$. The kinematics is determined by two parameters $x$ and
$\xi$ \cite{GKST83,TEL90,TEL95}. The first one
\begin{equation}
x=\frac{4E \omega_0}{ m^2c^4}= 0.019\left[\frac{E}{\GEV}\right]
\left[\frac{\MKM}{\lambda}\right]
\end{equation}
determines the maximum energy of the scattered photons:
$\omega_m=Ex/(x+1) \sim 4\gamma^2\omega_0 \;\;(x \ll 1)$.  If the
electron beam is cooled at the initial energy $E_0 = 5\; \GEV$ (after
damping ring and bunch compression) and $\lambda = 0.5\; \MKM$
(Nd:glass laser) then $x_0 \simeq 0.2$ (we will provide $E$ and $x$
with the index 0 for designation of their  values at the begining of
a cooling region). The second parameter is
\begin{equation}
\xi^2 = \left(\frac{eB_0\hbar}{m\omega_0 c}\right)^2,
\end{equation}
where $B_0$ is the  magnetic (or electric) field strength in the
laser wave.  At $\xi^{2} \ll 1$ an electron interacts with one
photon from the field (Compton scattering, undulator radiation), while
at $\xi^2 \gg 1$ an electron scatters on many laser photons
simultaneously (synchrotron radiation (SR), wiggler). We will see that
in the considered method $\xi^2$ may be ``small'' and ``large''.

In the cooling region near the laser focus the r.m.s radius of the
laser beam depends on the distance $z$ to the focus (along the beam)
in the following way \cite{GKST83}: $ r_\gamma = a_\gamma \sqrt{ 1 +
z^2 /Z_R^2}$, where $Z_R =2\pi a^2_\gamma /\lambda$ is the Rayleigh
length (an effective depth of laser focus), $a_\gamma$ is the
r.m.s. focal spot radius.  The density of laser photons is $n_\gamma =
(A/\pi r^2_\gamma \omega_0) \exp (-r^2/r^2_\gamma) F_\gamma(z+ct)$,
where $A$ is the laser flash energy and $\int F_\gamma(z) dz = 1.$

In the case of strong field ($\xi^2 \gg 1$) it is more appropriate to
speak in terms of strength of the electromagnetic field which is
$\bar{B^2}/4\pi=n_{\gamma}\omega_0, B=B_0cos(\omega_0t/\hbar-kz)$.
Assuming $F_{\gamma}=1/l_{\gamma}$ and $Z_R \ll l_{\gamma} \simeq l_e$
and using the classical formula for radiation loss
($dE/dx=(2/3)r^2_e\gamma^2 B^2,\;r_e=e^2/mc^2$) we obtain the ratio of
emittances before and after the laser target
\begin{equation}
 \frac{\EN_0}{\EN} \simeq \frac{E_0}{E} = 1+\frac{r_e^2}{3m^2c^4} \int
B_0^2dz = 1+\frac{64\pi^2r_e^2\gamma_0}{3mc^2\lambda\l_e}A \
\end{equation}
\begin{equation}
 A[J] = \frac{25\lambda[\MKM\ ]l_e[\MM\ ]}{E_0[\GEV\ ]}
\left(\frac{E_0}{E}-1\right).
\end{equation}
These equations are correct at $x \ll 1$ for any value
of $\xi^2$.  For example: at $\lambda=0.5\; \MKM,\; l_e=0.2\; \MM,
E_0=5\;\GEV, E_0/E=10$ the required laser flash energy $A = 4.5\; J.$
To reduce the laser flash energy in the case of long electron bunches,
one can compress the bunch (length) before cooling as much as possible
and stretch it after cooling up to the required value.

The eqs (3,4) were obtained for $Z_R \ll l_{\gamma} \sim
l_e$ and give the minimum flash energy for a certain ratio $E_0/E$.
To further estimate  the photon density at the laser focus we
will assume $Z_R \sim 0.25l_e$. In this case, the required
flash energy is still close to its minimum, but the field strength
is not so high as for very small $Z_R$. From the previous
equation for $Z_R = 0.25 l_e$ it follows $B_0^2/(8\pi) =
\omega_0n_{\gamma}=A/(\pi a_{\gamma}^2 l_e) =8A/(\lambda l_e^2)$.
Substituting $B_0$ into (2) we get
$$\xi^2 = \frac{16r_e\lambda A}{\pi l_e^2 mc^2} =
\frac{3\lambda^2}{4\pi^3 r_e l_e \gamma_0}\left(\frac{E_0}{E}-1\right
) \ = $$
\begin{equation}
\;\;\;\ = 4.3 \frac{\lambda^2 [\MKM\ ]}{l_e [\MM\ ]E_0 [\GEV\ ]}
\left(\frac{E_0}{E}-1\right ).
\end{equation}
Example: for $\lambda$ = 0.5 \MKM, $E_0$ = 5 \GEV\ , $ E_0/E$ = 10,
$l_e$ = 0.2 \MM\ (the NLC project) $\Rightarrow \xi^2$ = 9.7. For
larger bunch lengths and shorter wave lengths, $\xi^2$ may be smaller.
So, both ''undulator'' and ''wiggler'' cases are possible.

Later we will see that in order to have lower limit on emittance and
smaller depolarization it is necessary to have a low $\xi^2$. With a
usual optics one can reduce $\xi^2$ only by increasing $l_{\gamma}$
(and $Z_R$) with a simultaneous increase of the laser flash
energy. From (4) and (5) we get
\begin{equation}
A \propto \frac{\lambda^3}{\gamma_0^2 \xi^2}\left(\frac{E_0}{E}-1 \right)^2.
\end{equation}

Is it possible to reduce $\xi^2$ keeping all other parameters
(including flash energy) constant?  Yes, providing a way to stretch
the focus depth without changing the radius of this area is found. In
this case, the collision probability (or $\int B^2dz$) remains the same
but the maximum value of $\xi^2$ will be smaller.  A solution of this
problem was given in \cite{TEL96}. It is based on the
non-monochromaticity of the laser light and the chirped pulse
technique. In this scheme, the cooling region consists of many laser
focal points (continuously) and light comes to each point exactly at
the moment when the electron bunch is there.  One can consider that a
short electron bunch collides on its way sequentially with \n\
(``number of focuses'') short light pulses of  length $l_{\gamma}
\sim l_e$ and focused with $2Z_R \sim l_e$.  There is one
restriction on \n: along the cooling length $L \approx \n\cdot l_e$ the
transverse size of an electron beam should be smaller than the laser
spot size $a_{\gamma} \simeq \sqrt{\lambda Z_R/2\pi} \sim
\sqrt{\lambda l_e/4\pi}$.  In further examples we will use $\n\ \sim 10$
for stretching the cooling region from 100 \MKM\ to 1 \MM.

The electron energy spread   arises from the
quantum-statistical nature of radiation.  After energy loss
$\Delta E$, the increase of the energy spread
$\Delta(\sigma_E^2)=\int\varepsilon^2 \dot{n}(\omega) d\omega dt
=-aE^2\Delta E$,  where $\dot{n}(\omega)$ is the spectral density
of photons emitted per unit time, $a=14\omega_0/5m^2c^4=7x_0/10E_0$ for
the Compton case and $a=55\hbar e B_0/(8\pi\sqrt{3}m^3c^5)\;$ for the
``wiggler'' case \cite{WID}.

There is a second effect which leads to  decreasing  the energy
spread.  It is due to the fact that $dE/dx \propto E^2$ and an electron
with higher (lower) energy than the average  loses more (less) than on
average.  This results in the damping: $d(\sigma_E^2)/\sigma_E^2 =
4dE/E $ (here $dE$ has negative sign). The full
equation for the energy spread is $d\sigma_E^2 = -aE^2dE +
4(dE/E)\sigma_E^2$, with  solution
$$\frac{\sigma_E^2}{E^2} = \frac{\sigma_{E_0}^2E^2}{E_0^4}+
aE_0\frac{E}{E_0}\left(1-\frac{E}{E_0}\right)\sim$$
\begin{equation}
  \;\; \sim\frac{\sigma_{E_0}^2E^2}{E_0^4}+
  \frac{7}{10}x_0(1+\frac{275\sqrt{3}}{336\pi}\xi)\frac{E}{E_0}
  \left(1-\frac{E}{E_0}\right).
\end{equation}
Here the result for the Compton scattering and SR are joined together.
Example: at $\lambda=0.5\;\MKM,\; E_0=5\; \GEV\ (x_0 = 0.19)$ and
$E_0/E=10$,  the first Compton term alone gives $\sigma_E/E \sim 0.11$
and with the second term ($\xi^2 = 9.7$, see the example above)
$\sigma_E/E \sim 0.17$.

What $\sigma_E/E$ is acceptable? In the last example $\sigma_E/E \sim
0.17$ at E = 0.5 GeV. This means that at the  collider energy E =
250 GeV we will have $\sigma_E/E \sim 0.034\%$, that is better than
necessary (about 0.1 \%).

In a two stage cooling system, after reacceleration to the initial
energy $E_0$ = 5 GeV the energy spread is $\sigma_E/E_0 \sim
1.7\%$. For this value there may be a problem with focusing of
electrons which can be solved using a focusing scheme with
correction of chromatic aberations.  What are the resources if a
smaller energy spread is necessary?  The energy spread after
reacceleration is $\sigma_E/E_0 =(\sigma_E/E)(E/E_0)$. One can find
that the first (Compton) term $ \sigma_E/E_0 \propto
(E_0/\lambda)^{1/2}(E/E_0)^{3/2}$; the second (SR) $\propto
(E_0/l_e)^{1/4}(E/E_0)^{5/4}$ for $l_{\gamma} \sim l_e$ and $\propto
\lambda^{1/4}(E_0/E)/A^{1/4}$ for free $A$ (and $l_{\gamma} \sim Z_R >
l_e$).  Stretching of the cooling region also helps: $\sigma_E/E_0
\propto 1/\n^{1/4}$ (only the second term).

{\it Resume:} the energy spread in the one stage cooling scheme is not
a problem; for the multistage cooling system one has to use a
special focusing system with chromatic corrections in front of each
next stage.

{\it The minimum normalized emittance} is determined by the quantum nature
of the radiation. Let us start with the case of  pure Compton
scattering at $\xi^2 \ll 1$ and $x_0 \ll 1$. In this case, the
scattered photons have a uniform energy distribution: $dp = (3/2)[1-2\omega/\omega_m+2(\omega/\omega_m)^2]
d\omega/\omega_m$, where $\omega_m = 4\omega_0\gamma^2$. The angle of
the electron after scattering is \cite{GKST83} $\theta_1^2 =
(\omega_m\omega - \omega^2)/(\gamma^2E^2)$. After averaging over the
energy spectrum we get the average $\theta_1^2$ in one collision:
$\langle\theta_1^2\rangle = 12\omega_0^2/(5m^2c^4)$. After many Compton
collisions ($N_{coll}$) the r.m.s. angular spread in i=x,y projection
$\Delta\langle\theta^2_i\rangle = 0.5\Delta\langle\theta^2\rangle =
0.5N_{coll}\langle\theta_1^2\rangle = -0.5(\Delta
E/\bar{\omega})\langle\theta_1^2\rangle = -3\omega_0\Delta E/5E^2$.

The normalized emittance $\ENI{^2} = (E^2/m^2c^4)\langle r_i^2\rangle
\langle\theta_i^2\rangle$ does not change when $\Delta
\langle\theta_i^2\rangle/ \langle\theta_i^2\rangle$ = $-2\Delta E/E.$
Taking into account that
$\langle\theta_i^2\rangle\equiv\ENI\ /\gamma\beta_i$ we get the
equilibrium emittance due to the Compton scattering
$$   \ENI_{, min} \approx
 0.5\gamma E \beta_i\Delta \langle\theta_i^2\rangle/\Delta E =
 \frac{3\omega_0}{10mc^2}\beta_i
= \frac{3\pi}{5}\frac{\lambda_C}{\lambda}\beta_i =$$
\begin{equation}
  =\frac{7.2\cdot 10^{-10}\beta_i [mm]}{\lambda [\MKM\ ]} \;\;
  \mbox{m$\cdot$rad},
\end{equation}
where $\lambda_C=\hbar/mc=3.86\cdot10^{-11}$ \CM.  For example: $\lambda =0.5\; \MKM,
\;\beta=l_e/2=0.1\;\MM\ (NLC) \Rightarrow
\EN_{,min}=0.8\cdot10^{-10}\;\M\cdot$rad. For comparison in the NLC
project the damping rings have $\ENX\ =3\cdot 10^{-6}\; \M\cdot$rad,
$\ENY\ =3\cdot10^{-8}\; \M\cdot$rad.

Let us consider now the case $\xi^2 \gg 1$ when the electron moves as
in a  wiggler. Assume that the wiggler is planar and deflects the
electron in the horizontal plane. If an electron with  energy E
emits a photon with energy $\omega$ along its trajectory the
emittance changes as follows \cite{WID}: $\delta \EX =
(\omega^2/2E^2)H(s);\; H(s)=\beta_x\eta_x^{\prime 2}
+2\alpha_x\eta_x\eta_x^{\prime} + \gamma_x\eta_x^2$; where $\alpha_x =
-\beta_x^{\prime}/2,\; \gamma_x = (1+\alpha_x^2)/\beta_x,\; \beta_x $
is the horizontal beta-function, $\eta_x$ is the dispersion function,
$s$ is the coordinate along the trajectory. For $\beta_x = const$ the
second term in {\it H} is equal to zero, the second term in a wiggler
with $\lambda_w \ll \beta$ is small, so that {\it
H(s)} $\approx \beta\eta^{\prime 2}$. In a sinusoidal wiggler field
$B(z)=B_wcos\;k_wz,\; k_w=2\pi/\lambda_w$,
$\eta^{\prime\prime}=1/\rho$ ($\rho$ is the radius of curvature) one
finds that $\eta^{\prime} = (eB_w/k_wE)\; sin\;k_wz$.  The increase
of \EX\ on a distance $dz$ is
$$\Delta \E_x = \int \frac{H}{2}\left(\frac{\omega}{E}\right)^2
\dot{n}(\omega)d\omega dt = \frac{55}{48\sqrt{3}}\frac{r_e\hbar
c}{(mc^2)^6} E^5 \langle \frac{{\it H}}{\rho^3}\rangle dz,$$
where $ \langle {\it H}/\rho^3\rangle_w =
8\beta_x\lambda_w^2(eB_w)^5/(140E^5\pi^3)$ for the wiggler and
$\dot{n}(\omega)$ is the spectral density of photons emitted per unit
time.  The energy loss averaged over the wiggler period is $\Delta E =
r_e^2B_w^2E^2dz/(3m^2c^4)$. The normalized emittance $\EN=\gamma\E\ $
is not changed when $Ed\E\ + \E\ dE =0$. Using this
and replacing $B_w$ by 2$B_0$, $\lambda_w$ by $\lambda/2$ we obtain
the equilibrium normalized emittance in the linear polarized
electromagnatic wave for $\xi^2 \gg 1$
$$\ENX = \frac{11e^3\hbar c \lambda^2 B_0^3
\beta_x}{24\sqrt{3}\pi^3(mc^2)^4} =
\frac{11}{3\sqrt{3}}\frac{\lambda_C}{\lambda}\beta_x \xi^3 \approx$$
\begin{equation}
\approx \frac{8\cdot 10^{-10}\beta_x [\MM\ ] \xi^3}{\lambda [\MKM\ ]} \;\;
\mbox{m$\cdot$rad}.
\end{equation}
Using eq.(5) we can get a scaling of the minimum \ENX\ for a multistage
cooling system with a cooling factor $E_0/E$ in one stage: $\ENX\
\propto \beta_x\lambda^2(E_0/E)^{3/2}/(l_e\gamma_0)^{3/2}$ when
$l_{\gamma} \sim l_e$(minimum A) and $\ENX\ \propto
\beta_x\lambda^{7/2}(E_0/E)^{3}/(\gamma_0^3A^{3/2})$ for free A and
$\l_{\gamma} > l_e$ (for $\beta_x = const$).  Stretching the laser
focus depth by a factor \n\ , one can further reduce the horizontal
normalized emittance: $\ENX\ \propto 1/\n^{1/2}$(if $\beta_x \propto
\n$).  For our previous example we have $\xi^2 =9.7$ and $\ENX\ =
5\cdot 10^{-9}\;\M\cdot$rad (in the NLC \ENX\ = 3$\cdot 10^{-6}$
m$\cdot$rad). Stretching  the cooling region with \n=10, further
decreases the horizontal emittance by a factor 3.2.

Comparing with the Compton case (8) we see that in the strong
field the horizontal emittance is larger by a factor
$\xi^3$. The origin of this factor is clear: $\ENX \propto
\eta_x^{\prime 2} \omega_{crit.}$, where $\eta_x^{\prime} \sim
\xi\theta_{compt.}$ and $\omega_{crit.} \sim \xi\omega_{compt.}$.

Let us  roughly estimate the minimum vertical normalized emittance at
$\xi \gg\ 1$. Assuming that all photons are emitted at an angle
$\theta_y = 1/(\sqrt{2}\gamma)$ with the $\omega = \omega_c$ similarly
to the Compton case, one  gets $\Delta\langle\theta^2_y\rangle =
(\omega_c \Delta E)/(2\gamma^2E^2) = -(3e\hbar \bar{B_w}\Delta
E)/(4E^2mc)$.  Using the first part of eq.(8) we get
$$\ENY_{min} \sim \frac{3}{8}\frac{\hbar e \bar{B_w}}{m^2c^3}\beta_y =
\frac{3}{2\pi}\frac{\hbar e \bar{B_0}}{m^2c^3}\beta_y = 3
\left(\frac{\lambda_C}{\lambda}\right)\beta_y\xi \approx$$
\begin{equation}
\approx \frac{1.2\cdot 10^{-9}\beta_y [\MM\ ] \xi}{\lambda [\MKM\ ]} \;\;
\mbox{m$\cdot$rad}.
\end{equation}
For the previous example (NLC beams), eq.(10) gives $\ENY_{min} \sim
7.5\cdot 10^{-10}$ m$\cdot$rad (for  comparison in the NLC project
$\ENY=3\cdot10^{-8}$ m$\cdot$rad). The scaling: $\ENY\ \propto
\beta_y(E_0/E)^{1/2}/(l_e\gamma_0)^{1/2}$ when $l_{\gamma} \sim
l_e$(minimum A) and $\ENY\ \propto
\beta_y\lambda^{1/2}(E_0/E)/(\gamma_0A^{1/2})$ for free A and
$\l_{\gamma}>l_e.$

For arbitrary $\xi$ the minimum emittances can be estimated as the sum
of (8) and (9) for \ENX\ and sum of (8) and (10) for \ENY\
\begin{equation}
  \ENX \approx
  \frac{3\pi}{5}\frac{\lambda_C}{\lambda}\beta_x(1+1.1\xi^3);\;\; \ENY
  \sim \frac{3\pi}{5}\frac{\lambda_C}{\lambda}\beta_y(1+1.6\xi).
\end{equation}

Finally, let us consider the problem of the depolarization. For the
Compton scattering the probability of spin flip in one collision is
$w=(3/40)x^2$ for $x \ll 1$ (it follows from formulae of
ref.\cite{KOT}). The average energy losses in one collision are
$\bar{\omega} = 0.5xE$. The decrease  of polarization degree
 after many collisions is $dp = 2wdE/\bar{\omega} = (3/10)x(dE/E) =
(3/10)x_0(dE/E_0)$.  After integration, we obtain the relative decrease
of the longitudinal polarization $\zeta$ during one stage of the  cooling
(at $E_0/E \gg 1$)
\begin{equation}
        \Delta\zeta/\zeta = 0.3x_0\;\;\; \propto E_0/\lambda,
\end{equation}
For $\lambda = 0.5\; \MKM\ $ and $E_0 = 5\; \GEV\ $, we have $x_0 =
0.19$ and $\Delta\zeta/\zeta = 5.7 \%$. This is valid only for $\xi^2
\ll 1.$

In the case of strong field ($\xi^2 \gg 1$) the spin flip probability
per unit time is the same as in the uniform magnetic field \cite{LAN}
$w = (35\sqrt{3}r_e^3\gamma^2ce\bar{B^3})/(144\alpha(mc^2)^2)$,
where for the wiggler $\bar{B^3} = (4/3\pi)B_w^3$. Using the relation
between $dE$ and $dt$ in the wiggler we get
\begin{equation}
  \frac{\Delta\zeta}{\zeta} =
  \int\frac{35\sqrt{3}er_eB_0}{9\pi\alpha(mc^2)^2}dE \sim
  \frac{35\sqrt{3}}{36\pi}x_0\xi.
\end{equation}
For the general case, the depolarization can be estimated as the sum of
equations (12) and (13)
\begin{equation}
 \Delta\zeta/\zeta \approx 0.3x_0(1+1.8\xi).
\end{equation}
For the previous example with $\xi^2 =9.7$ and $x_0 = 0.19$ we get
$\Delta\zeta/\zeta = 0.057+0.32 = 0.38$, that is not acceptable. This
example shows that the depolarization effect imposes the most
demanding requirements on the parameters of the cooling system. The main
contribution to depolarization  gives the second term. One can decrease
$\xi$ by increasing $l_{\gamma}$ and $Z_R$. In this method,
the required flash energy increases and the attainable $\xi$ depends on the
available laser flash energy. From (14) and (6) we can get a scaling for
the second term $ \Delta\zeta/\zeta \propto
\lambda^{1/2}(E_0/E-1)/A^{1/2}.$ Another method is  stretching of the
focus depth, which does not require  increasing  laser flash energy.
Stretching by a factor \n\ reduces the second term as $1/\sqrt{\n}$.
After stretching the cooling region by a factor \n=10 we get
$\Delta\zeta/\zeta = 0.057+0.1\sim 15\%$.

{\it Conclusion.} Possible sets of parameters for the laser
cooling: $E_0 = 4.5$ GeV, $l_e=0.2 $ mm, $\lambda = 0.5$ \MKM, flash
energy $A \sim$ 5--10  J, focusing system with stretching factor
\n=10. The final electron bunch will have an energy of 0.45 \GEV\ with an
energy spread $\sigma_E/E \sim 13 \%$, the normalized emittances
\ENX, \ENY\ are reduced by a factor 10, the limit on the final emittance is
$\ENX\ \sim\ENY\ \sim2\cdot10^{-9}\;$ m$\cdot$rad at $\beta_i= 1\;
\MM$, depolarization $\Delta\zeta/\zeta \sim 15\%$.  If the focus
depth stretching technique works, we can hope on further reduction of
depolarization.  The two stage system with the same parameters gives
100 times reduction of emittances (with the same restrictions).  The
maximum emittance at the entrance (the electron beam radius is two
times smaller than the laser spot size) is about $10^{-5} $m$\cdot$rad
(the increase of this number is possible after some optimization).

For the cooling of the electron bunch train one laser pulse can be
used many times.  According to (3) $\Delta E/E = \Delta A/A$ and even
25\% attenuation of laser power leads only to small additional energy
spread.

The proposed scheme of laser cooling of electron beams seems  very
promising for  future linear colliders and allows to reach ultimate
luminosities. It is useful especially for photon colliders, where
collision effects allow considerable a increase of the luminosity.
Perhaps this method can be used for  X-ray FEL based on high energy
linear colliders.

\vspace*{0.7cm} I would like to thank Z.Parza, the organizer of the
Program ''New Ideas for Particle Accelerator'' at ITP, UCSB, Santa
Barbara, supported with National Science Foundation Grant NO
PHY94--07194.  I am grateful to A.~Skrinsky for  very useful
discussions and critical remarks concerning polarization. Also I would
like to thank D.~Cline, S.~Drell, I.~Ginzburg, J.~Irving, G.~Kotkin,
D.~Leith, B.~Palmer, A.~Sessler, V.~Serbo, B.~Richter, P.~Zenkevitch
for useful discussions.

\vspace*{0.5cm}


\begin{thebibliography}{99}
%
\bibitem{LOW} Low et al., International Linear Collider Technical Review Committee Report No. SLAC-Rep-471, 1996.
%
\bibitem{GKST81} I.F.~Ginzburg, G.L.~Kotkin, V.G.~Serbo, V.I.~Telnov,  Pisma Zh. Exsp. Teor. F
{\bf 34}, 514 (1981) [JETP Lett. {\bf 34} 491 (1982)].
%
\bibitem{GKST83} I.F.~Ginzburg, G.L.~Kotkin, V.G.~Serbo, V.I.~Telnov, Nucl. Instr.
 Meth. Phys. Res., {\bf 205}, 47 (1983).
%
\bibitem{TEL90} V.I.~Telnov, Nucl. Instr. Meth.  A {\bf 294},
72 (1990).
%
\bibitem{TEL95} V.Telnov,
 Nucl.Instr. Meth. Phys. Res. A {\bf 355}, 3 (1995).
%
%
\bibitem{BERK} {\it Proc.of Workshop on \GG\ Colliders,
Berkeley CA, USA, 1994}, Nucl. Instr. Meth. Phys. Res. A
 {\bf 355}, 1--194 (1995).
%
\bibitem{TEL96}V.Telnov, {\it Talk at the Int.Symp. New Modes of
Part. Accel. Techn.  Sources}, ITP, UCSB, Santa  Barbara, August
19--23, 1996, NSF-ITP-96-142, SLAC-PUB-7337.
%
\bibitem{WID} H.Wiedemann, {\it Particle Acc. Physics: basic principles
    and linear beam dinamics}, Springer-Verlag, 1993.
%
\bibitem{TRAV} C.Travier, Nucl.Instr. Meth. A {\bf 340}, 26 (1994).
%
\bibitem{palmer} R.Palmer, in Ref.~[6], p.150
%
\bibitem{LAN} V.Berestetskii, E.Lifshitz and L.Pitaevskii, {\it Quantum
Electrodynamic}, Pergamont press, Oxford, 1982.
%
\bibitem{KOT} G.Kotkin, S.Polityko, V.Serbo, Yad. Fiz. 
 {\bf 59}, 2229 (1996).
\end{thebibliography}
\end{document}